\documentclass[12pt,a4paper,twocolumn]{narms2}
\usepackage{amsmath}
\usepackage{graphicx}
\newcommand{\be}{\begin{equation}}
\newcommand{\ee}{\end{equation}}
\newcommand{\bi}{\begin{itemize}}
\newcommand{\ei}{\end{itemize}}

\usepackage{subfigure} 
\usepackage{psfig}
\usepackage{timesmt}   
\usepackage{chikako}   

\makeatletter
\renewcommand{\section}{\@startsection%
{section}%
{1}%
{0mm}%
{- \baselineskip}%
{0.15\baselineskip}%
{\normalfont\normalsize}}%

\renewcommand{\subsection}{\@startsection
{subsection}%
{2}%
{0mm}%
{-\baselineskip}%
{0.15\baselineskip}%
{\normalfont\normalsize}}%
\makeatother

\title{Sound wave velocities in dry and lubricated granular packings packings: 
numerical simulations and experiments}
\author{
{\large I. Agnolin, J.-N. Roux} \\
{\em Laboratoire des Mat\'eriaux et des Structures du G\'enie Civil, Institut Navier, 
Champs-sur-Marne, France}\\
\\
{\large P. Massaad, X. Jia, P. Mills} \\
{\em Laboratoire des Milieux Divis\'es et des Interfaces,
  Universit\'e de Marne-la-Vall\'ee, Champs-sur-Marne, France} 
}
\date{}
\abstract{ABSTRACT: numerical simulations are used to investigate the origins of the 
different wave velocities measured in dense granular samples assembled with different methods.
Glass bead packings are prepared in the lab either by pouring and vibrating 
the dry material in a container, or by mixing with a very small amount of a viscous lubricant. Lubricated samples,
although less dense, exhibit significantly higher wave velocities for confining pressures in the 100~kPa range. 
Numerical predictions for
elastic moduli agree much better with experimental results when the computational preparation
of the samples mimics the laboratory one, albeit in a simplified manner. A plausible explanation to the
laboratory observations is that the coordination number, 
which influences the material stiffness more than its density, is notably higher in lubricated packings.}
\begin{document}
\maketitle
\frenchspacing  
\section{INTRODUCTION}
The mechanical properties of granular packings are sensitive to fine geometric details
of the particle arrangement, as very small motions can modify the
force-carrying contact network. In practice, direct measurements of such important internal state
variables as the coordination number and the distribution
of contact orientations~\shortcite{RR04} are usually impossible.

In this context, ultrasonic wave propagation
measurements might provide useful information on the internal
structure of granular samples under confining stresses. 
Wave propagation has been used as a probe to investigate the
microstructure of granular packings by several physics
groups~\shortcite{jia99,JM01,gilles03}. 
At low frequencies such that the wavelengths are very long compared to the heterogeneity 
of the medium, the granular medium is effectively a homogeneous continuum to the propagating wave, 
while at high frequencies when the wavelength decreases down to the order of the grain size, 
scattering effects caused by the spatial fluctuations of force chains lead to diffusive transport of sound waves \shortcite{Jia04}. 

Wave propagation has also become a standard method to measure elastic moduli in geotechnics
laboratories~\shortcite{GBDC03}, where rheological
testing devices are often equipped with specially designed transducers~\shortcite{LIGR01}. 
The recent soil mechanics
literature~\shortcite{THHR90,GBDC03} made it clear that
``dynamic'' measurements of elastic moduli (wave
propagation or resonance modes) agree with ``static''
ones (slopes of stress-strain curves), provided
strain increments are small enough (below $10^{-5}$).
Correctly measured elastic moduli therefore determine long wavelength
sound velocities in granular materials as in ordinary solids. 

Discrete particle simulations can also be used to evaluate elastic
properties of model granular
materials~\shortcite{JNR97a,Makse04,JNR04}. If experimental
packings are correctly simulated, such
studies can clarify the relations between wave propagation
measurements and sample microstructure. 

We first briefly report here (Sec.~\ref{sec:exp}) on experimental measurements on
pressure-dependent sound velocities and attenuation in dense samples
of dry glass beads (hereafter denoted as E1 samples), as well as lubricated ones in which the effects
of intergranular friction are strongly reduced in the preparation
stage. The resulting materials (denoted as E2 samples) are, remarkably, less dense, but stiffer 
(with notably larger elastic wave velocities).
 
In order to investigate the microscopic origins of these results, 
we use numerical simulations, as presented in Sec.~\ref{sec:num}, which is the main part of the present communication. We resort to
suitable, yet simplified, models to mimic laboratory assembling
procedures (thus producing sets of samples denoted as A and B by a lubricated procedure, and a third series
C by a ``vibrated'' procedure, see Sec.~\ref{subsec:numprep} for details). 
Comparisons of numerical results to experimental ones lead to 
an interpretation of the observed differences between dry
and lubricated samples in terms of coordination number.

\section{EXPERIMENTS\label{sec:exp}}
\begin{figure}[!htb]
\centering
\includegraphics[height=12cm]{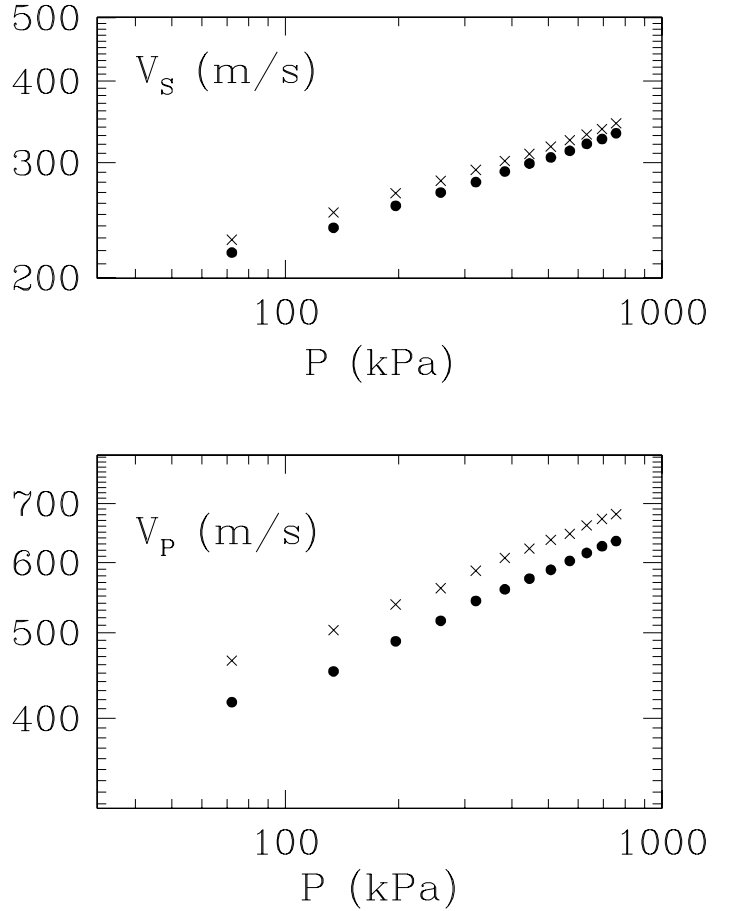}
\caption{Velocity of longitudinal (bottom plot) and transverse (top
  plot) sound waves as a function of confining pressure for
laboratory samples E1 (dots, assembled by vibrating dry grains) and E2 (crosses, lubricated). 
\label{fig:vppexp}
}
\end{figure}

The glass beads used in our experiments are of diameter d = 0.3-0.4
mm, randomly deposited by pouring and vibrating in a duralumin
cylinder of diameter W = 30 mm and varying height from 10 mm to 30~mm.
The container is closed with two fitting pistons and a normal load
is applied to the granular sample across the top piston. Before the
ultrasonic measurements, one cycle of loading-unloading is performed
in the granular packing in order to consolidate the sample and
minimize its hysteretic behavior. A plane-wave generating transducer
of diameter 30 mm (top piston) and a detecting transducer of diameter
30~mm are placed on the axis at the top and bottom of the cylindrical
container in direct contact with the glass beads. Once the cell is filled,
a normal load corresponding to apparent pressures P ranging from 30
kPa to 1000 kPa is applied to the upper piston using a jackscrew
arrangement, while the lower piston is held fixed, as in oedometric
loading. (The preloading pressure cycle reached up to 400~kPa).
Broadband short pulse excitations of $2\,\mu$s duration
centered at a frequency of 500~kHz are applied to the source
transducer. The time of flight of the transmitted ultrasonic signal is used
to measure the sound speeds, which are shown on figure~\ref{fig:vppexp}, and to which results
of numerical simulations will be compared in Sec.~\ref{subsec:numvp} (Figure~\ref{fig:vpp}). 
Measurements are first performed on assemblies of dry beads, initially prepared in a dense state 
by layerwise deposition and tapping (samples E1).
Then, to study the influence of
intergranular friction on the initial structure and elastic properties of the
medium, we mix a small amount of liquid lubricant (trioleine, volume fraction
0.5\%) with the granular sample for tens of minutes to distribute the
oil uniformly among the grains. 
We observe (figure~\ref{fig:vppexp}), that the solid fraction in the obtained 
lubricated samples (denoted as E2) 
is about $0.62$, lower than in the dense dry ones E1
( $\sim 0.64$), while sound wave speeds are significantly higher for
E2 specimens (by 10 to 20\%) than in the E1 case. 
It is also apparent on Figure~\ref{fig:vppexp} (a logarithmic plot), that
the increase of sound velocities with pressure is slightly faster in E1 samples.
\section{NUMERICAL SIMULATIONS \label{sec:num}}
We now turn to numerical simulations to explore the origins, at the scale of the contact network,
of the differences between the dry (E1) and the lubricated (E2) samples.
\subsection{{\em Sample preparation and characterization.}\label{subsec:numprep}}
Our numerical simulations, like others~\shortcite{Makse04}
(but, admittedly, unlike the experiments of Sec~. II)
focus on homogeneous, \emph{isotropic} states, under varying
pressure $P$. The samples -- the same as those studied
by~\citeN{Ivana-ici}-- comprise 4000 identical beads
of diameter $a$ in a periodic cell which changes size as stresses are applied. 
Averages and standard deviations are
evaluated on 5 such samples. 

Spherical grains are attributed the elastic properties of glass beads 
(Young modulus $E=70$~GPa, Poisson coefficient $\nu=0.3$), and
assembled by a standard molecular dynamics method. The
contact laws involve Coulomb friction and a suitably simplified form
of (Hertz-Mindlin) contact elasticity
(see~\shortcite{Ivana-ici} and references therein). 

The most
frequently used numerical procedure~\shortcite{Makse04} to obtain
dense configurations is to suppress friction altogether while
compressing a granular gas to mechanical equilibrium under a given
isotropic pressure, thus simulating \emph{perfect lubrication}. 
This results in typical isotropic random close
packing structures~\shortcite{OSLN03}, the properties of which are not sensitive to the
details of the procedure, provided it is fast enough to bypass
crystal nucleation entirely. On applying this method with prescribed pressure
$P=10$~kPa, we obtained samples we denote as A. Their
coordination number $z^*$, evaluated on eliminating inactive grains
(``rattlers'') from the count, is close to 6, the value it should approach~\shortcite{JNR2000,Makse04}
in the rigid limit, as the average contact deflection $h$ becomes negligible,
$h/a \propto (P/E)^{2/3} \to 0$. On imposing larger pressure levels
(up to quite high values in simulations) a friction coefficient $\mu=0.3$ is
introduced. This implicitly assumes that the perfect lubrication of the
assembling stage (in which intergranular forces are transmitted through a
very thin layer of lubricant) disappears as large
static pressures bring solid surfaces into contact.
Another series of numerical samples, type B ones, are also made
assuming only a very low  friction in the initial stage ($\mu_0=0.02$),
as a first model of imperfect lubrication. Finally, in order to
imitate the preparation of dense dry samples of type E1 in the laboratory, 
a third series C of initial states under 10~kPa is prepared
as follows. First, A configurations are slightly dilated, scaling
coordinates by a common factor $\lambda=1.005$; then grains are mixed
at constant volume, as by thermal agitation~: the system is
thus strongly shaken in a dense state; the final step is a compression,
with friction ($\mu=0.3$) and some viscous dissipation to
mechanical equilibrium at $P=10$~kPa. 

\begin{figure}[!htb]
 \centering
 \subfigure[$\Phi$ versus $P$.]{
  \includegraphics[width=4.2cm]{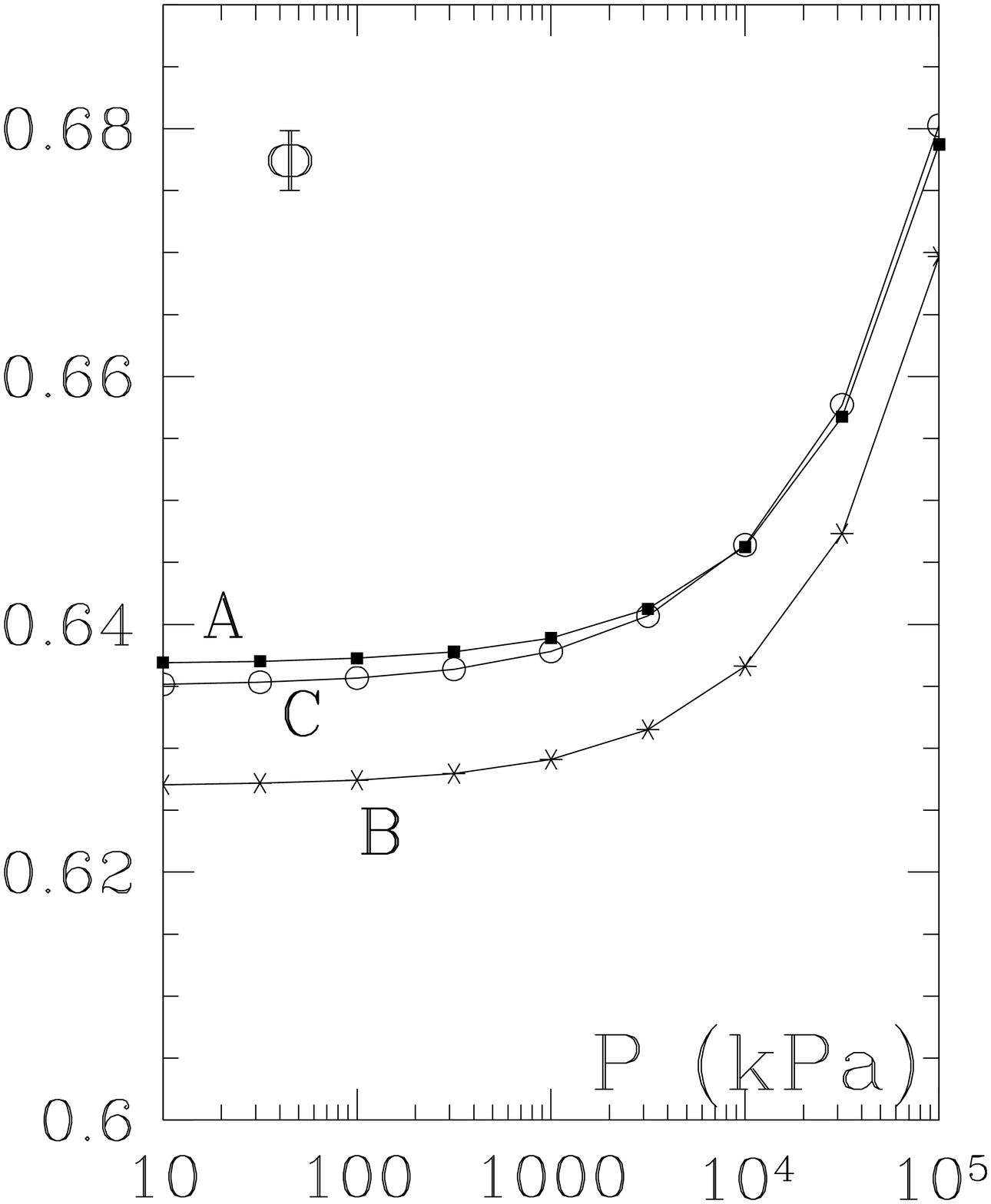}
  \label{fig:phip}
 }
 \subfigure[$z^*$ versus $P$.]{
  \includegraphics[width=4.2cm]{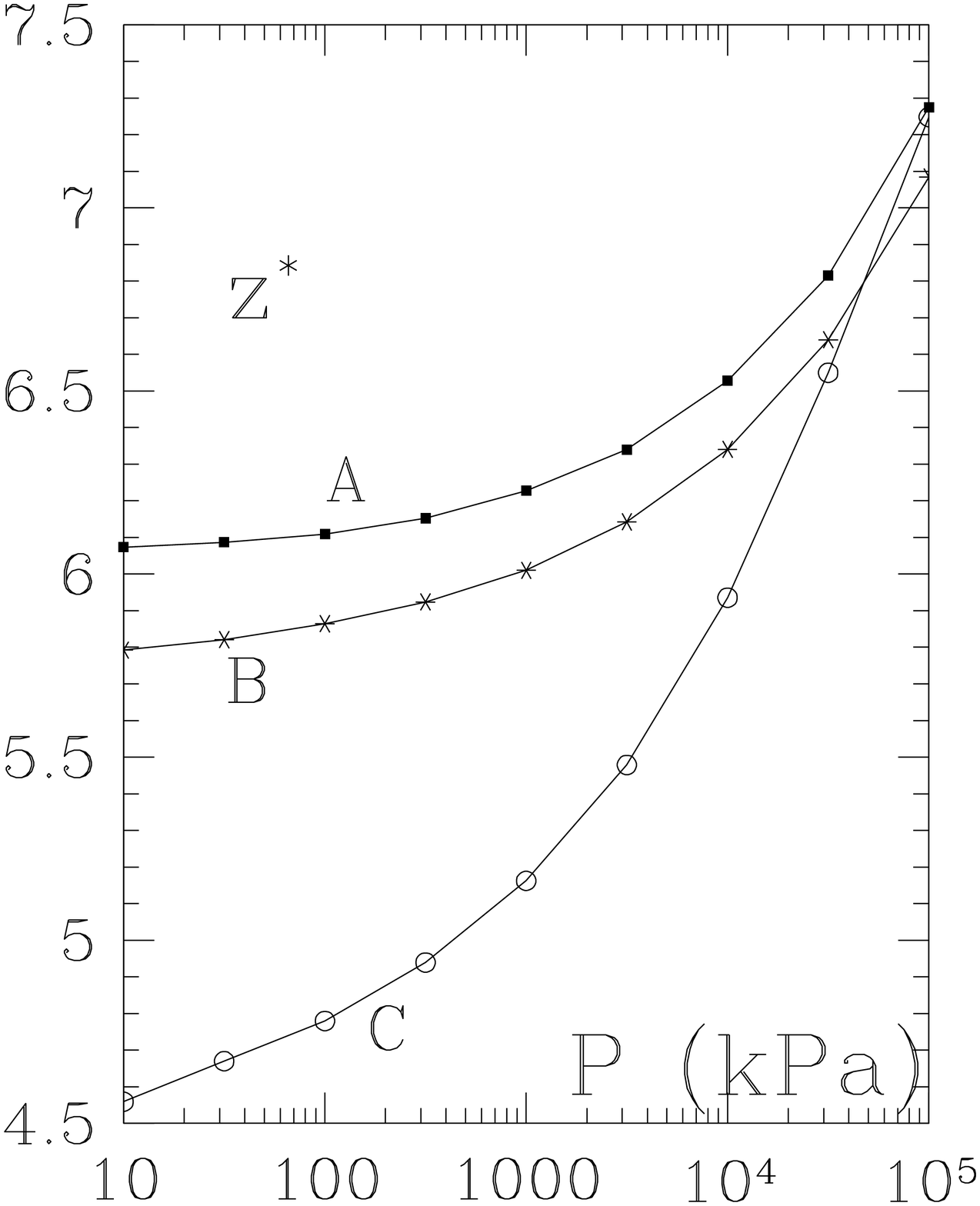}
  \label{fig:zp}
 }
 \caption[]{Volume
   fractions and coordination numbers as functions of $P$ in
   isotropically compressed samples of types A (square dots), B
   (stars), and C (open circles).
\label{fig:zphip}
}
\end{figure}

Figure~\ref{fig:zphip} displays the values of solid volume fraction
$\Phi$ and coordination number $z^*$ as functions of pressure $P$ in
states A, B and C.
While B samples are similar to A ones, with somewhat lower values
of $\Phi$ and $z^*$, C configurations, remarkably, 
although very nearly as dense as A ones (as expected given
their preparation method), and actually denser than B ones, exhibit much lower coordination numbers.
The difference between C and A states is gradually  reduced as $P$ grows.
\subsection{{\em Wave velocities.}\label{subsec:numvp}}
Isotropic states A, B, C possess two independent elastic constants, the bulk ($B$)
and shear ($G$) moduli, from which the velocities of longitudinal
($V_P$) and transverse ($V_S$) sound for large wavelengths are deduced
as ${\displaystyle V_P = \sqrt{\frac{B+4G/3}{\rho}}}$ and ${\displaystyle V_S =
\sqrt{\frac{G}{\rho}}}$,
$\rho $ denoting the mass density in the granular material. 

\begin{figure}[!htb]
\centering
 \includegraphics[width=9cm]{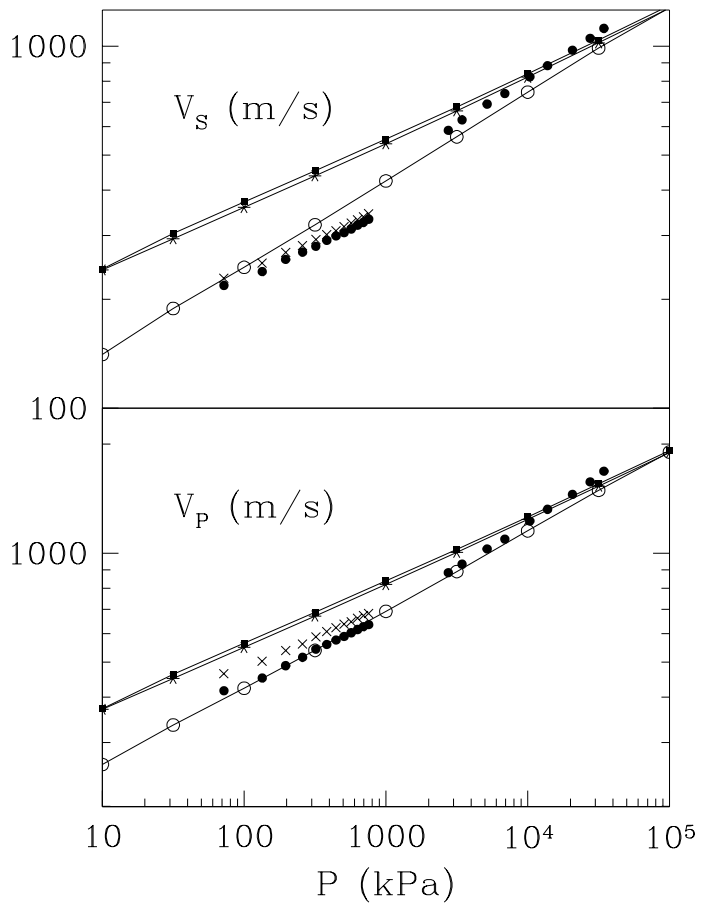}
\caption{
Comparisons between numerical and experimental values for $V_P$
(bottom plot) and $V_S$ (top part). Numerical results are plotted as
connected symbols, as on figure~\ref{fig:zphip}: square dots for A
(perfect lubrication), stars for B (imperfect lubrication),
open circles for C (``vibrated'' dry grains).
Experimental results are shown as filled
circles (E1, dry samples) and crosses (E2, lubricated samples) for $P$ in the
70~kPa-1~MPa range, as on Figure~\ref{fig:vppexp}, while the data obtained by Domenico on dry
glass beads are plotted as black dots for $P$ around 10~MPa.
\label{fig:vpp}}
\end{figure}

Once mechanical equilibrium states are obtained with sufficient
accuracy, the stiffness matrix of the contact
network is built, and both elastic moduli are obtained on solving
a linear system of equations, in which particle displacements and
rotations as well as strain increments are the unknown, and the
imposed stress increments determine the right-hand
side. Figure~\ref{fig:vpp} presents the simulated values of $V_P$ and
$V_S$ for all three sample types and compares them to experimental results.
Scarcely coordinated C samples are clearly in much better agreement with experimental data on
dense, dry packings than A or B ones in the $P\sim 100$~kPa range. The difference between B and C systems 
is qualitatively the same as the one between dry and lubricated laboratory packings: a lower
density, but higher wave velocities that increase a little slower with
pressure. Some indications about the pressure depence of elastic
moduli $B$ and $G$ in A and C configurations, and a discussion of
their prediction by micromechanical modeling schemes, are given by
\shortciteN{Ivana-ici} in these proceedings.
Despite the crudeness of the numerical models for ``shaking'' (C) or
``lubrication'' (A, B), they appear to capture the right experimental trends. We suggest therefore that the
larger stiffness (or sound speed) in lubricated packings is due to their larger coordination number.

We therefore conclude that C-type samples are better models for dense dry
granular packings (of type E1). Such a statement apparently contradicts the good agreement 
reported by \shortciteN{Makse04} between numerical measurements of sound speeds on samples of type A  
and experiments on dry specimen of type E1, such as those by \shortciteN{DO77}, shown on Figure~\ref{fig:vpp}.
However, on confronting simulation
results with experimental ones, those authors focussed on a higher pressure
range (above several MPa, see Figure~\ref{fig:vpp}), in which A and
C-type systems are much less differentiated. We checked that our
numerical results on A samples coincide nearly perfectly with those of
\shortciteN{Makse04}. Thanks to our experimental results for smaller confining pressures we can also distinguish between
dense systems with high and low coordination numbers, the latter ones being more appropriate as models for dense, dry 
granular assemblies.

It might also be noted that in a different communication in the present
proceedings~\cite{JN-ici}, it is argued that the mechanical properties
of C-type samples in quasistatic, axisymmetric triaxial compression
are closer to those observed in the
laboratory with sand or glass bead samples assembled by pouring and
shaking than those of A configurations.
\section{CONCLUSION}
Numerical simulations   
show that wave propagation in dense bead packings assembled with various procedures can reveal
differences in their microstructure, beyond the sole density. 
Admittedly, more accurate comparisons with
experiments are certainly necessary. This requires a more detailed knowledge of laboratory samples and their
anisotropic state of stress \shortcite{Yac-ici}, a possible consideration of the effects
of capillary adhesion, as well as more realistic numerical models for
laboratory procedures \shortcite{Sacha-ici}.
However, the results presented here show that discrete numerical simulations can relate
experimentally accessible data on wave speeds to internal variables
such as coordination number or fabric. 
They clearly stress the need for a better understanding of the influence of
the preparation procedure on the subsequent mechanical properties in
quasistatic conditions. They suggest a plausible interpretation of the
observed larger wave velocity in lubricated systems.
Numerical procedures, even for isotropic samples, should not be
selected as appropriate because they produce the right
density. Coordination number, a largely independent state variable,
strongly influences elastic properties and quasistatic stress-strain
laws~\cite{JN-ici}. 

\end{document}